
\input phyzzx.tex
\PHYSREV
\overfullrule0pt
\sequentialequations
\def\etal{{\it et al.}}
\def\to{\rightarrow}

\FIG\figA{%
A typical two-gluon exchange contribution to the amplitude
for $\gamma^*(q) p \to V(q+\Delta) (p-\Delta)$.}%
\FIG\figB{%
Light-cone perturbation theory graphs for
the scattering of a quark-antiquark pair by a colorless target.}

\Pubnum={SLAC--PUB--6412R\cr
CU--TP--617 \&
UCD--93--36}
 \date{January 1994}
 \pubtype{(T/E)}
 \titlepage
{\tenpoint \baselineskip 12.5pt
 \title{DIFFRACTIVE LEPTOPRODUCTION
OF VECTOR MESONS IN QCD\foot{%
Work supported by the Department of Energy, contracts
DE--AC03--76SF00515, DE-FG02-93ER40771, and DE--FG03--91ER40674,
Binational Science Foundation Grant 9200125, and
Texas National Research Laboratory grant RGFY93--330.}}
 \author{Stanley J. Brodsky}
 \SLAC
 \andauthor{L. Frankfurt\foot{On leave of absence from the
 St. Petersburg Nuclear Physics Institute, Russia.}}
 \address{School of Physics and Astronomy\break
Raymond and Beverly Sackler Faculty of Exact Sciences\break
Tel Aviv University, Ramat Aviv 69978, Israel}
 \andauthor{J. F. Gunion}
\address{Davis Institute for High Energy Physics\break
Department  of  Physics,  University  of California \break
Davis,  California  95616}
\andauthor{A. H. Mueller}
\address{Department  of  Physics, Columbia University,
 New York, New York 10027}
\andauthor{M. Strikman\foot{Also at  the St. Petersburg Nuclear
Physics Institute, Russia.}}
\address{Department of Physics, Pennsylvania State University,
University Park, Pennsylvania 16802}
\par}
\vfill
\endpage
\vglue 1in

\centerline{\bf Abstract}
\smallskip
We demonstrate that the distinctive features
of the forward differential cross
section of diffractive
leptoproduction of a vector meson
can be legitimately calculated in
perturbative QCD in terms of
the light-cone $q \bar q$ wave function of the vector meson
and the gluon distribution of the target. In particular,
we calculate the $Q^2$ and nuclear dependence of the  diffractive
leptoproduction of vector mesons and estimate the cross section.
The production of longitudinally polarized vector
mesons by longitudinally polarized virtual photons is predicted to
be the dominant component, yielding a cross section behaving as $Q^{-6}$.
The nuclear dependence of the diffractive
cross sections, which follows from a
factorization theorem in perturbative QCD,
provides important tests of color transparency as well as
constraints on the shadowing of the gluon
structure functions and the longitudinal structure functions of nuclei.

\vfill
\submit{Physical Review D.}
\vfill
\endpage

\chapter{Introduction}

We shall consider in this paper the small momentum transfer coherent
electroproduction of vector mesons, $\gamma^*(q) + p \to V(q + \Delta) +
(p - \Delta),$ where the target,
$p$, can be either a nucleon or a nucleus, and where the state
$(p-\Delta)$ is identical
to $p$ except for a small momentum transfer.  Here $V$ can be any
possible vector meson,
 $\rho^0,\omega, \phi, J/\psi, \Upsilon.$ We shall be concerned with the
kinematic region where $x = Q^2/s,$ and $M_V^2/s$  are small while
$Q^2/\Lambda_{QCD}^2$ is large.  Small $x$
means that large longitudinal distances, in the rest frame of the
target, are involved.  The
effective longitudinal distance during which the process takes place is
large: $\Delta z \approx {1\over M} {s \over Q^2}$ [1],
where $M$  is the target mass and $s = (p + q)^2.$
Our interest is in the possibility of applying perturbative QCD (PQCD)
to the calculation of hard processes characterized by
large longitudinal distances.
We demonstrate an interesting interplay
of perturbative and nonperturbative
QCD effects  in the region where the coupling constant is small
but distances are large which leads to a new way
to probe light-cone wave functions of hadrons.
In the case where $\Delta z \gg 2R,$ with $R$
the radius of a nuclear target, we predict
that interesting color transparency
effects will occur in diffractive electroproduction of vector mesons.
In the last section of the paper, we  briefly discuss
QCD predictions for the nuclear dependence of the diffractive cross
sections and show how such measurements
can provide important constraints on the shadowing of the gluon
structure functions and of the deep-inelastic
longitudinal structure functions of nuclei.

In general, the physics underlying our PQCD
calculation can be directly tested through
the striking nuclear effects predicted in vector meson leptoproduction
which differ from those that would result
within the Glauber approximation.
However, the thrust of the present paper
is not a detailed discussion of nuclear effects, but rather a general
description of the $Q^2$-dependence of the forward differential cross
section, an analysis of the rigor of the PQCD calculation,
a semi-quantitative estimate of the magnitude of the cross
section, a prediction as to which polarization
dominates the large $Q^2$ reaction, and a calculation of the nuclear
dependence of electroproduction of vector mesons.

The main features of our perturbative QCD analysis agree with those
obtained in the non-perturbative approach proposed by Landshoff and
Donnachie [2] and discussed in more detail by Cudell [3]. In the DL model, the
Pomeron is represented by the effective exchange of two non-perturbative
gluons coupling via an effective constant coupling  and the
$\rho$ wavefunction is approximated as a non-relativistic vertex where
the quark and anti-quark have equal four-momenta.  In our analysis, the
two gluon aspect of the QCD Pomeron emerges automatically at large $Q^2$
in a form directly related to the proton's gluon structure function.  In
addition we treat the relativistic structure of the vector meson
generally.  We find that the leading twist contribution to $\rho$
leptoproduction is controlled by the $\rho$ distribution amplitude
$\phi(z,Q),$ the valence $q \bar q $ wavefunction which controls large
momentum transfer exclusive processes.  As in the DL model, PQCD predicts
that the dominant leptoproduction amplitude couples a longitudinal photon
to a longitudinally polarized vector meson, and the leading cross section
$\sigma_L(s,Q^2)$ falls as $1/Q^6.$ The agreement of the data with this
form is shown in Cudell's paper.[3]

We also predict that  the
cross section for production of a transversely polarized vector meson
will fall as $Q^{-8}$.
In fact, it is the
end-point contributions (which complicate the analysis
of elastic processes) that yield the dominant contribution
to the cross section for a transversely polarized $V$. In contrast,
we show that end-point configurations are unimportant for
the diffractive production of longitudinally polarized vector mesons.

Our work is also closely related to that
of Ryskin [4] who has made detailed
calculations of $J/\psi$ electroproduction in leading-logarithm in
PQCD, employing the constituent quark model.
In this work we focus on the
dominant longitudinally polarized channels and find
that an analogous analysis can be  applied  to light vector mesons.
However,  if in our formulae we take a nonrelativistic approximation
for the wave function of the vector meson, then we find that
the cross section for the production of the
$J/\psi$ at large $Q^2$  is a factor of 4
less than that calculated in Ref. [4]
in the nonrelativistic approximation.
Unlike the heavy quarkonium case where the amplitude is controlled by the
wavefunction at the origin and the lepton pair decay constant,  light
hadron leptoproduction probes the shape of the minimal Fock-state
wavefunctions, \ie\ the hadron distribution amplitudes $\phi_V(z,Q)$
as defined in Ref. [5].
This dependence allows tests of non-perturbative predictions from QCD sum
rules and lattice gauge theory. We also find that this
sensitivity may help explain the pattern
of $SU(3)$-flavor symmetry breaking
seen in the leptoproduction data.

The electroproduction of vector
mesons has also been recently discussed by
Kopeliovich et al. [6] (KNNZ) in the context of their constituent
quark model approach to high energy, small momentum transfer
processes  in which the constituent quarks interact
perturbatively with the nucleon.
Indeed, there is much in common between the KNNZ approach and the
present discussion, including their use of
Eq. (20) for a nucleon target as derived and applied
in Refs. [7] and [8].
However,  the application of the nonrelativistic quark model to
hard processes is
questionable for this application, since in QCD
hard processes should be calculated through the distribution
of current, not constituent quarks [5].  Moreover, we find that
the application of PQCD is only legitimate
for the production of longitudinally polarized vector mesons.
Thus, in distinction to Ref. [6],
we expect a different  nuclear dependence  for
the production of transversely and longitudinally
polarized vector mesons. Further, the
eikonal
approximation
used by KNNZ for the nuclear case is at variance with
Eqs.~(33) and (37), which are derived from
factorization in PQCD. As a result we expect that nuclear
effects will be leading twist---logarithmically decreasing at large $Q^2$
and increasing  at small $x$.
In contrast,  KNNZ suggest, on the basis of their constituent quark
model,  that the nuclear effects in
diffractive leptoproduction are a higher
twist effect.

In order to achieve a simple result for the forward differential cross
section, see \eg\ Eq.~(34),
we find it necessary to work in the leading $\ell n \
1/x$ and the leading $\ell n \ Q^2/\Lambda_{QCD}^2$ limit.
As discussed in the body of the paper we believe
that it is also possible
to calculate $d\sigma/dt$ at small\ $t$\ in the leading $\ell n\ 1/x$
approximation in terms of the unintegrated gluon distribution discussed
sometime ago by Catani, Ciafaloni
and Hautmann [9], without using the leading $\ell n\ Q^2/\Lambda_{QCD}^2$
approximation.  However,
such a calculation is not likely to lead to such a simple result
as Eq.~(34), which follows
in the leading-double-logarithmic approximation.
Thus, we have not pursued the single-logarithmic
calculation, although it is clearly interesting to do so.
At the same time we will explain at the end of Section 2.2
that corrections to the expression we obtain are numerically small.

Our final results should be taken with some caution.
Since they are based on a
leading-logarithmic, even leading-double-logarithmic, calculation, the
normalizations may not be
completely reliable.  However, the $Q^{-6}$ dependence, the dominance
of longitudinal polarization for both
the virtual photon and the produced
vector meson, and the proportionality of
the cross section to the unintegrated
gluon distribution in the target, and
therefore the nontrivial dependence of
cross section on atomic number, are firm predictions which
should not depend on our logarithmic approximations.

\chapter{The Diffractive Cross Section in QCD Near t=0.}

In this section the near-forward differential cross section for
 $\gamma^*(q) + p \to V(q + \Delta) + (p - \Delta)$ will be calculated in
QCD.  The target, labelled by its momentum $p$, is scattered into a state
of momentum $p - \Delta$ which we assume to be a
particle of the same species as $p$.
For example, $p$ and $p -\Delta$ may
both refer to  protons.  $V$ is a vector meson of mass $m_V.$
We assume $s/m_V^2 \gg 1,\ s/Q^2\gg 1$ and
$- t \equiv\ -\ \Delta^2 \ll Q^2$, and $s = (p+q)^2, q^2 =\ -\ Q^2.$
We also suppose $Q^2/\Lambda_{QCD}^2$,
and $Q^2/m_V^2$ are both much greater than one.

The differential cross section for the process described above is
$${d\sigma_{\lambda\lambda^\prime}\over dt} = {1\over 16\pi s^2}
\bigl\vert {\cal M}_{\lambda\lambda^\prime}\bigr\vert^2\eqn\sigdef$$
where $\lambda$ is the polarization of the virtual photon and
$\lambda^\prime$ is the polarization of the final state vector meson.
In the large $s$ and large $Q^2$ limit, but with $s/Q^2 \gg 1,$
we expect the amplitude ${\cal M}$ to be dominated by two-gluon
exchange, a particular graph of which is illustrated in Fig. \figA.
We will prove this statement for the production of
longitudinally polarized vector meson.  In Fig. \figA, lines $k^\prime$
and $q-k^\prime$ are the quark and antiquark making up the vector meson.

\section{
Setting\ up\ the\ Calculation\ of\ the\ Matrix\ Element}

In proceeding to calculate ${\cal M}$ it is
useful to view the process in a
physical  way.  To that end, we choose a frame
where $p$ is essentially at rest
(\ie, $p_+ \ll q_+$) and where
$$q = (q_+, q_-, q_\perp) = \left(q_+, {-Q^2\over q_+}, 0\right)
\eqn\framedef$$
with $q_+= q_0 + q_3, q_- = q_0-q_3.$ Then $(q + \Delta)^2 =m_V^2$
and $(p - \Delta)^2 = M^2$ give
$$\Delta_- \approx {(Q^2 + m_V^2 + \Delta_\perp^2)\over q_+},\qquad
\Delta_+ \approx -
p_+ {\Delta_\perp^2\over M^2}.\eqn\deltaforms$$
The polarization vectors are
$$\epsilon^\gamma(\perp) = (0,0,\epsilon_\perp^\gamma),\
\epsilon^\gamma(L) = \left({q_+\over Q},\ { Q\over q_+},\ 0_\perp\right)
\eqn\polgam$$
for the virtual photon and
$$\epsilon^V(\perp) = (0,0,\epsilon_\perp^V),\ \epsilon^V(L) =
\left({(q_++\Delta_+)\over m_V},
\ {-m_V\over (q_++\Delta_+)},\ 0_\perp\right)
\eqn\polv$$
where we have dropped terms proportional to $\Delta_\perp$ in
\polv.

The process illustrated in Fig. \figA\
takes place, sequentially in time, as
follows.
\item{(i)} The virtual photon breaks up into a quark-antiquark pair
with a lifetime $\tau_i$ given by
$$q_+ \tau_i^{-1} = Q^2 + {k_\perp^2+ m^2 \over z(1-z)}
\approx Q^2\ .\eqn\tauinitial$$
Here $m$ is the current quark mass.
This estimate is valid for the production of a longitudinally polarized
vector meson only. In the case of a transversely polarized
vector meson,
the end-point non-perturbative contribution arises from the kinematical
region where $z$ is close to 0 or 1: $z,1-z \sim  m^2/ Q^2$
as in the aligned-jet model of Ref. [10] adapted to QCD in Ref. [11].
(See also the discussion below.)
\item{(ii)} The quark-antiquark pair then scatters off the target proton.
\item{(iii)} The quark-antiquark pair
then lives a time $\tau_f$ determined by
$$q_+ \tau_f^{-1} = {k_\perp^2+m^2\over z(1-z)}\eqn\taufinal$$
before the final state vector meson is formed.  We note that
$\tau_f \geq \tau_i.$

\noindent Thus, the amplitude ${\cal M}$ can be written as a
product of three factors:
(i) the wavefunction giving the amplitude for the virtual photon to break
into a quark-antiquark pair;
(ii) the scattering amplitude of the
quark-antiquark pair on the target; and
(iii) the wavefunction giving the amplitude for the scattered
quark-antiquark pair of flavor\ $f$\ to become a vector meson.
Following the conventions of Ref.~[5], we have
$$\eqalign{
{\cal M}_f&={\sqrt{N_c}} \sum_{\lambda_1,\lambda_2}
\int
 {d^2k_\perp d^2k^\prime_\perp\over (16\pi^3)^2}\ \int_0^1 dz
 \int_0^1 dz^\prime  \cr\crr &\quad \times
\psi_{\lambda_1\lambda_2}^{V*}(k_\perp^\prime,z^\prime)
T_{\lambda_1\lambda_2}(k_\perp^\prime,z^\prime;  k_\perp,z)
\psi_{\lambda_1\lambda_2}^\gamma (k_\perp,z)\ ,\cr}
\eqn\mform$$
where $\lambda_1$ and $\lambda_2$ are the helicities of the
quark-antiquark pair which are conserved during the scattering off the
target. $\psi^V$ and $\psi^\gamma$ are the light-cone wavefunctions
in the notation of Ref. [5].
We have explicitly extracted the sum over the
$N_c$ colors of the quarks and the $1/\sqrt{N_c}$ from the color
singlet normalization of $\psi^V$.  Thus, $\psi^V$ and $T$ correspond
to the wave function and scattering amplitude for a quark-antiquark
pair of definite color.

We only obtain a simple result, in terms of the gluon distribution of
the proton, in the leading-logarithmic approximation (in longitudinal and
transverse momentum) for $T$.
In this leading-logarithmic approximation, the
time of scattering of the quark-antiquark pair with the
target is much less than $\tau_i$
so that $T$ is
effectively
given
as the {\it on-shell} scattering of
a quark-antiquark pair off the target.
Our task now is to evaluate $\psi^\gamma$
and $T$ and thus to express the amplitude, ${\cal M}_f,$
in terms of integrals over the exclusive wavefunction $\psi^V.$

\section{%
$\psi^\gamma$ and
the Scattering of the Quark-Antiquark Pair by the Target}

The evaluation of $\psi^\gamma$ at lowest order in $e$ is
straightforward.  In the convention of Ref.~[5],
$$\psi_{\lambda_1\lambda_2}^\gamma(k_\perp,z) =
e e_f {\overline u_{\lambda_1}(k)
\gamma \cdot \epsilon^\gamma v_{\lambda_2}(q-k)\over {\sqrt{k_+}}
\left(q_- -
{k_\perp^2+m^2\over k_+} - {k_\perp^2+m^2 \over (q-k)_+}\right)
{\sqrt{(q-k)_+}}}\ ,\eqn\psigammaform$$
which leads to
$$\psi_{\lambda_1\lambda_2}^\gamma(k_\perp,z) =\ -\ e e_f\
{\overline u_{\lambda_1}(k)
 \gamma \cdot \epsilon^\gamma v_{\lambda_2}(q-k)\over {\sqrt{z(1-z)}}
\left(Q^2 + {k_\perp^2+m^2 \over z(1-z)}\right)}\ .\eqn\psigamgeneral$$
In Eq.~\psigamgeneral, $e$ is the
charge of the proton and $e_f$ is the charge
of a quark of flavor $f$, as a fraction of the proton's charge.  We omit
the label $f$ on $\psi^\gamma$ for simplicity of notation.
We see from \polgam\ that longitudinal polarization
apparently gives the dominant contribution, though we must wait until we
have calculated $\psi^V$ to see that this is indeed the case.
Anticipating this result we set $\epsilon^\gamma =
\epsilon^\gamma (L)$ and, using \polgam, obtain
$$\psi_{\lambda_1\lambda_2}^\gamma (k_\perp,z) =\ -\ {e e_f
Q\delta_{\lambda_1,-\lambda_2}\over Q^2 + {k_\perp^2+m^2\over
z(1-z)}}\ .\eqn\psigam$$

Now consider the scattering of the quark-antiquark pair by the target.
The relevant light-cone perturbation theory
graphs are shown in Fig.~\figB.
(It is important to note our convention of using the momenta $q-k^\prime$
and $k^\prime+\Delta$ that enter $\psi^V$
to define the momenta of the lines.)
As we have mentioned before, a simple result emerges only in the
leading-logarithmic approximation in $\ell n \ Q^2/s.$
In that case, $\ell_+/k_+ \ll 1$
and the dominant couplings of the lines $\ell$ and
$\ell + \Delta$ to the quark-antiquark pair occur with a $\gamma_+$
(in a covariant gauge).
Thus, the vertices of the lines $\ell + \Delta$ and $\ell$ with the
quark-antiquark pair are exactly the same for each of the graphs in
Fig.~\figB.
Further, all energy denominators are dominated by the $\ell$ and
$\ell + \Delta$ lines so that the energy denominators are also the same
for each of the graphs in Fig.~\figB.
The differences between the different graphs in
Fig.~\figB\  are only in the
labelling of the momenta on the left-hand side
of the diagram as they emerge from the $\psi^\gamma$ wavefunction.
The result obtained in the $\Delta_\perp\ll \ell_\perp$ limit being
considered is very simple:
$$\eqalign{
T_{\lambda_1\lambda_2}&(k_\perp^\prime,z^\prime;k_\perp,z)= \cr
& \times 16\pi^3
\int \bigl\{2\delta(k^\prime_\perp -k_\perp) - \delta(k^\prime_\perp -
k_\perp + \ell_\perp)- \delta(k^\prime_\perp
- k_\perp - \ell_\perp) \bigr\}\cr
&\hbox{\hskip .50in}\times \delta(z^\prime-z) {\cal I}(\ell)
 {d^2\ell_\perp d\ell_+\over 16\pi^3}\ ,\cr
}\eqn\tform$$
%
where ${\cal I}(\ell)$ represents the gluon propagators and zero-angle
gluon-nucleon scattering amplitude shown in Fig. 1.

Before attempting to evaluate the $d^2\ell_\perp d\ell_+$ integral
involving ${\cal I}$,  let us first do the
integrals over $d^2k^\prime_\perp dz^\prime / 16\pi^3$
indicated in \mform.  Using \tform\ one arrives at the combination
$$2\psi_{\lambda_1\lambda_2}^\gamma(k^\prime_\perp,z) -
 \psi_{\lambda_1\lambda_2}^\gamma (k^\prime_\perp + \ell_\perp,z) -
\psi_{\lambda_1\lambda_2}^\gamma (k^\prime_\perp - \ell_\perp,z) = \Delta
\psi_{\lambda_1\lambda_2}^\gamma\eqn\psigamdifcomp$$
which, using \psigam, gives
$$\Delta \psi_{\lambda_1\lambda_2}^\gamma =
{-2 e e_f \delta_{\lambda_1,-\lambda_2}  Q \ell_\perp^2\over
\left[Q^2 + {k^{\prime\ 2}_\perp +m^2\over z(1-z)}\right]^2
z(1-z)}\ ,\eqn\psigamdif$$
when $\ell_\perp^2/Q^2 \ll 1.$
Thus (relabelling $k^\prime_\perp\to k_\perp$),
$${\cal M}_f = {\sqrt{N_c}}\ \Sigma_{\lambda_1,\lambda_2} \int
 {d^2k_\perp\over 16\pi^3} \int_0^1 dz\,
\psi_{\lambda_1\lambda_2}^{V*}(k_\perp,z)\  {\cal I}(\ell)\
 {d^2\ell_\perp d\ell_+\over 16\pi^3} \ \Delta
\psi^\gamma_{\lambda_1\lambda_2}\, .\eqn\mformii$$

If instead of a final state vector meson we were considering a
virtual photon identical to the initial state photon, the exact same
dependence on
$\ell_\perp$ and $\ell_+$ would appear.
This allows one to identify an integral over ${\cal I}(\ell) $ with the gluon
distribution.
Indeed, in the
leading-logarithmic approximation in $\ell n \ 1/x$ and $\ell n \
Q^2/\Lambda^2,$
where ${\cal I}$ is purely imaginary,
$$\int\ {d^2\ell_\perp d\ell_+\over 16\pi^3} \
\ell_\perp^2 {\cal I}(\ell) =
i {4\pi^2T_R\alpha_s\over N_c}\ (s+Q^2)\, x G(x,Q^2)\ ,\eqn\igrelation$$
where the factor of $T_R/N_c$ arises by virtue of averaging over the
color and matching anti-color of the initial quark and antiquark.
(In the usual convention $T_R = \half.$)
The simplest way to determine the normalizing factor  in Eq. (2.16)
is to compare the integral on the left hand side of (2.16) with the known
relation of the longitudinal structure function with the gluon density.
Once one
is sure that the gluon distribution should emerge from the integration on the
left-hand side of \igrelation, the normalization is most
easily set by taking the target to be
a quark and calculating the lowest order
contribution to {\cal I}. However, to establish that $x=Q^2/s$ and
$Q^2$ are the appropriate arguments in $xG(x,Q^2)$ requires employing the
leading $\ell n\ 1/x$ and $\ell n\ Q^2/\Lambda_{QCD}^2$ approximations.

We can understand these values from the
simple picture on which Eqs. \mform\ and \mformii\ are based.
Namely, we must have a strict sequence of events in which
(i) the virtual photon breaks up into an approximately on-shell
quark-antiquark pair; followed by (ii)
the scattering of the on-shell quark-antiquark pair by the
target; and, finally,
(iii) the scattered quark-antiquark pair turns into the vector meson.
This picture requires that
the time of scattering of the quark-antiquark
pair must be much less than $\tau_i$
and $\tau_f$ defined earlier: \ie\ $\tau \ll \tau_i \leq \tau_f$.
Using (recall that we employ OFPT here)
$$\tau \sim {1\over \ell_-} = {\ell_+\over \ell_\perp^2}\ ,\eqn\tauform$$
this condition reduces to
$$\ell_+\ll {q_+\ell^2_\perp\over Q^2}\,.\eqn\conditioni$$
Meanwhile,  the dominance of the imaginary part implies that
the $-$ components of the four-momenta
are approximately conserved from the initial to the intermediate
state where the $p+\ell$ line is cut, leading to $(p+\ell)_-\simeq p_-$
(neglecting $-$ components of order $1/q_+$). With this, we compute
$\hat s=(p+\ell)^2\simeq M^2+p_-\ell_+\ll\ell^2_\perp s/Q^2$, with the
last inequality coming from employing \conditioni.
In the leading $\ell n\ 1/x$ approximation we allow integration over
the parton-gluon scattering subprocess energy, $\hat s$,
up to this upper limit.  Recalling that the $x_g$
argument of $G(x_g,Q^2_g)$
is set by $x_g\hat s\sim \ell^2_\perp$, we see that in the
 leading $\ell n\ 1/x$
approximation $x_g$ should be identified with $x=Q^2/s$. To determine
the appropriate argument $Q^2_g$, we note that
the simple form for $\Delta \psi^\gamma$ given in \psigamdif\
depends on
$${\ell_\perp^2\over Q^2} \ll 1\ .\eqn\conditionii$$
In leading $\ell n \  Q^2/\Lambda_{QCD}^2$ we integrate $\ell^2_\perp$ up
to the maximum allowed
by this relation, and it is this maximum which determines the argument
$Q^2_g$ appearing in $G$.  The result is obviously $Q^2_g\sim Q^2$.
(We note here that so long as $\Delta_\perp^2/\Lambda_{QCD}^2 \ll 1,$
the left-hand side of  Eq. \igrelation\ does not depend
on $\Delta.$ From \deltaforms, we see that $\Delta_-$ is comparable to
$\ell_- = \ell_\perp^2/\ell_+$
only when $\ell_\perp^2/\ell_+$ reaches its smallest value, $Q^2/q_+
\approx \Delta_-.)$
It is useful to note that these conclusions match closely those obtained
in Ref. [7] where the scattering of a quark-antiquark pair
off a target is given by
$$\sigma(b^2)={2\pi^2 \over 3}\left(b^2 \alpha_s(Q^2)\bar x
  G_{N}(\bar x, Q^2)\right)_{\bar x=1/s{b^2},Q^2=1/b^2}\ ,
  \eqn\sigmaqqbar$$
where, for simplicity, the result has been stated for
the case $x_q \sim x_{ \bar q} \sim 1/2$.

Can one do better than the leading double log approximation? The answer
should be yes.
One should be able to eliminate the restriction
\conditionii\ and derive an
expression for ${\cal M}$ in
terms of integrals over the unintegrated gluon distribution and the
vector meson exclusive wavefunction.  The result, however, will be
significantly more complicated than the answer we are about to give for
the leading double-logarithmic approximation.
At the same time, the  corrections to \igrelation\ resulting from the
elimination of the
restriction \conditionii\ are numerically small.  A simple way to
justify this statement is to perform the
calculation in impact parameter space where
non-$\ell n\ Q^2/\Lambda_{QCD}^2$
corrections arise from the decomposition of
the matrix element of the real
part of $e^{i \ell_\perp \cdot b_\perp}-1 $  between
the wave functions of the virtual photon
$\gamma^*$ and  a vector meson in impact parameter space.
Here $b_\perp$ is the inter-quark distance in the meson wavefunction
and
$\ell_\perp$ is the transverse momentum of the gluon shown in
Fig.~\figB.
The second order term in $(\ell_\perp\cdot b_\perp)$ leads to
Eq.~\mformii. The corrections in question arise starting at the fourth
order in the expansion contain
a factor of $1/4!$ and are therefore small.

\section{%
The Final State Vector Meson Wavefunction}

Using \igrelation\ and \psigamdif\ in \mformii, and $(s+Q^2)\simeq s$,
one obtains
$${\cal M}_f = {-8is\pi^2 e e_f\alpha_s\over Q^3}\ {T_R\over\sqrt{N_c}}
\ x G(x,Q^2) \int
{d^2k_\perp\over 16\pi^3} \int_0^1 {dz\over z(1-z)}\ \Sigma_\lambda\
\psi_{\lambda-\lambda}^V (k_\perp,z)\eqn\mformiii$$
where
$$\psi^V_{\lambda-\lambda}(k_\perp,z)
= N_V{\overline v_{-\lambda}(q-k)\over
 {\sqrt{1-z}}}\ \gamma \cdot \epsilon^V {u_\lambda(k)\over {\sqrt{z}}}
\ \psi^V(k_\perp,z)\ {1\over m_V}\
.\eqn\psivform$$
Here, we have introduced a normalization factor $N_V$
for the particular spin state of $V$
being considered. (In the conventions
of Ref.~[5], $N_V=1/\sqrt 2$ for $\epsilon_V(L)$.)

We are now in a position to verify that longitudinal polarizations
give the dominant contribution.  From
\psigamgeneral\ and \psivform\ it is
straightforward to see that
longitudinal polarizations give the
leading contribution so long as $Q$ is
much larger than all quark and vector meson masses.  In this case,
using \polv\ in \psivform\ gives
$$\psi_{\lambda-\lambda}^V(k_\perp,z)
= \ - N_V \psi^V(k_\perp,z)\eqn\psiv$$
which yields (after performing $\Sigma_\lambda$ and setting $T_R=1/2$)
$${\cal M}_f = {8is\pi^2 e e_f \alpha\over Q^3{\sqrt{N_c}}}\ x G(x,Q)
N_V \int_0^1 {dz\over z(1-z)} \phi^V(Q,z)\eqn\mformiv$$
where
$$\phi^V(Q,z) = \int {d^2k_\perp\over 16\pi^3} \
\psi^V(k_\perp,z)\eqn\phiv$$
is the  distribution amplitude for longitudinally
polarized vector mesons
with the restriction $k_\perp^2 < Q^2$ understood for the
integration in \phiv. In getting the correct normalization for
the cross section, it will be important to note
that this result does not yet include the flavor
normalization for the $V$ wavefunction. (The spin normalization
factor is contained in $N_V$.)

In order to obtain an absolute normalization for ${\cal M}_f$, it is
necessary to relate $\phi^V$ to the experimental observable $f_V$
defined by
$$ \langle 0\vert J_{e.m.}^{\mu}\vert V\rangle=
{\sqrt 2 e f_V\over m_V \epsilon^{\mu}}\ ,\eqn\fvdef$$
in terms of which the decay width for $V\to e^+e^-$ is given by
$$\Gamma_V = {8\pi \alpha^2  f_V^2 
\over 3 m_V}\ .\eqn\gammav$$
We demonstrate in the Appendix
that
$$\int\, dz\,\phi^V(z)={f_V\over \sqrt{N_c}e_f2\sqrt2
N_V}\,.\eqn\phivintegral$$
Defining $\eta_V$ as
$$\eta_V\equiv\half
{\int\,{ dz \over z(1-z)} \phi^V(z)
\over \int\, dz \,\phi^V(z)}\,\eqn\etavdef$$
we obtain
$${\cal M}_f = {8\ i s\pi^2 f_V e\alpha_s \eta_V \over \sqrt 2 Q^3
N_c}\, x G(x,Q)\,.\eqn\mformv$$
The parameter $\eta_V$ is the effective inverse moment of the vector
meson distribution amplitude that controls the leading twist contribution
to the leptoproduction amplitude.  Higher particle number Fock state
amplitudes such as the $q \bar q g$ in a physical gauge
have a suppressed coupling to the
small-size quark pair;
in order to compensate their higher mass dimensions,
they must be accompanied
by further powers of $1/Q$.
Note that both $e_f$ and $N_V$ have cancelled out in relating
${\cal M}_f$ to the experimental observable $f_V$.

In order to estimate the cross section it is convenient to
consider two extreme examples for the shape of the $\rho$ distribution
amplitude:
$$\phi_1^V = {3\over \sqrt 2 N_V{\sqrt{N_c}}}\ {f_V\over e_f}
z(1-z)\eqn\phivasymptotic$$
$$\phi_2^V ={15\over \sqrt 2 N_V{\sqrt{N_c}}}\
{f_V\over e_f}
z(1-z)(1-2z)^2\ .\eqn\phivcz$$
Here $\phi_1$ is the asymptotic form of the
distribution amplitude, while $\phi_2$ is of the form suggested by
Chernyak and Zhitnitsky [12] for pions. The normalizations of $\phi_1^V$
and $\phi_2^V$ are chosen for consistency with Eq.~\phivintegral.
A simple calculation gives
$\eta_V=3$ for $\phi_1$ and
$\eta_V=5$ for $\phi_2.$
We  emphasize that in QCD amplitudes for hard
processes  are expressed
in terms of the minimal Fock component of the light-cone wave function of
the meson (see \eg\ Eq.~\mformiii), not in terms of constituent quark
model components. This is important when the
physics of color screening, which is relevant
to the transition from the non-perturbative to the perturbative regime,
is accounted for (\cf\ the discussion in Section 2.6
regarding the difference between the
production of transverse and longitudinally
polarized vector mesons, and the
discussion of color transparency in Section 2.8).

\section{%
The Flavor Dependence of the Cross Section}

Thus far, we have not discussed
the flavor dependence of the wavefunction of
the final state vector meson.  If
$V$  refers to a neutral $\rho$ meson, then the relevant wavefunction is
${1\over {\sqrt{2}}}(\vert u\bar u> - \vert d \bar d >).$
This would imply a replacement of $e_f$ in our calculations
of {\it both} ${\cal M}_f$ and $f_V$
by $e_f \to {1\over {\sqrt{2}}} (e_u-e_d) = 1/{\sqrt{2}}.$
However, since $e_f$ cancels out when ${\cal M}_f$ is
expressed in terms of $f_V$, as in  Eq.~\mformv,
the result is that we may use Eq.~\mformv\ for the
total amplitude without change provided we employ the appropriate value
for $f_\rho$. With our normalization conventions,
$f_\rho \approx 107$ MeV for the $\rho^0$.
If  $V$  refers to a $J/\psi$ one can simply use $e_f = e_c =
2/3$. Again, the explicit value of $e_f$ disappears if
${\cal M}_f$ is expressed in terms of $f_{J/\psi}$.
However, for the $J/\psi$ the value of $\eta_V$ is expected to
differ substantially from that for the $\rho$ since neither of the
wavefunctions given in \phivasymptotic\ and \phivcz\ is appropriate;
a wavefunction having $z\approx 1/2$ would be more suitable.

\section{%
The Differential Cross Section}

We shall now write the differential cross section for $\rho^0$
production.  Modifications for other lepto-produced neutral vector mesons
are straightforward and involve choosing
a distribution amplitude to replace
\phivasymptotic\ or \phivcz.
Using   Eq.~\mformv\ in  Eq.~\sigdef\  gives
$${d\sigma\over dt}\bigg\vert_{t=0} (\gamma^* N \to V N)
= {
8\pi^4 f_V^2\alpha_{EM}\alpha_s^2(Q)
\eta_V^2[xG(x,Q)]^2\over Q^6 N_c^2}\ . \eqn\dsigdt$$
It is important to keep in mind that
this result gives the differential cross section
for a longitudinally polarized photon to produce a
longitudinally polarized $\rho^0$; \ie\ it is not spin-averaged
over initial photon states.
As an aside, we note that
the $t$-dependence of this diffractive cross section is controlled by the
quasi-local two-gluon matrix element of the nucleon; in principle,
it could have a different fall-off than the elastic form
factors since the momentum transfer is shared by the two gluons.

It is convenient to rewrite Eq.~ \dsigdt\
in terms of the  leptonic width,
$\Gamma_V$
using Eq.~\gammav,
$${d\sigma\over dt}\bigg\vert_{t=0} (\gamma^* N \to V N)
= { 3 \pi^3 \Gamma_V m_V
\alpha_s^2(Q) \eta_V^2[xG(x,Q)]^2
\over \alpha_{EM} Q^6 N_c^2}\ \eqn\dsigdtt$$
since the coherent sum over the contributing flavors is identical for the
diffractive amplitude and the decay amplitude.
Our prediction  for the
$J/\psi$ leptoproduction cross section
using Eq.~\dsigdtt\ is smaller than
the one obtained in Ref.~[4] by a factor of 4, if
we assume, as in Ref.~[4],
a nonrelativistic form $\sim \delta(z-1/2)$ for the  $J/\psi$ meson wave
function so that $\eta_\psi \approx 2 $.


We can also use dispersion relations to determine the real part of the
amplitude.  For small enough $x$ and large $Q^2$ the contribution of the
real part is not negligible since the effective QCD pomeron intercept is
above 1.  Including the real contribution
as a perturbation
we get

$$ {d\sigma^{L}_{\gamma^* N \rightarrow V N} \over dt}\bigg\vert_{t=0} =
{ 3 \pi^3 \Gamma_V m_V \alpha_s^2(Q) \eta_V^2\mid (1+i{\pi\over 2}
{d\over d~ln~x}) x G_T(x,Q)\mid^2 \over \alpha_{EM} Q^6 N_c^2}.
$$

Since the gluon density rises rapidly at small $x$ in the large $Q^2$
domain, we predict a very substantial rise of the diffractive cross
section with energy at large $Q^2.$ For example, for $Q^2 \sim 10$
GeV$^2,$ we
predict a rise of the diffractive cross section at small $x$ by as much
as a factor of $100$ at HERA energies as compared to the cross section
measured at CERN.  Obviously this effect would also substantially modify
the $Q^2-$dependence of the cross section at large energies.

Let us now compare Eq.~\dsigdt\ with experimental data
for $\rho$-meson production at large $Q^2$ and small $x$.
Such data are available from EMC [13] and NMC [14,15]
and also from the E-665 FNAL experiment [16].
We will use the latest data of NMC [15]
which were obtained at the highest $Q^2$ with special attention
to removing backgrounds due to inelastic processes.
We also note that the data of
E-665 [16], which extend to somewhat smaller $Q^2,$
are generally consistent with NMC data.
All of these data confirm the important role of the longitudinal
contribution at large $Q^2$ from  the  measurement of the polarization of
the $\rho.$
If $s$-channel helicity conservation is assumed in the transition
$\gamma^* \rightarrow \rho,$ this  also determines
the ratio of $\sigma_L / \sigma_T$.

To convert the leptoproduction cross section $ d^2\sigma /dQ^2 d\nu$
from $\mu N$ to the virtual photoproduction cross section, we use
the standard relation
$\sigma _{\mu N}(Q^2,\nu ) = \Gamma \sigma _{\gamma ^*N}(Q^2,\nu ),$
where
$\Gamma  = \alpha_{EM} (\nu -Q^2/(2M))/(2\pi Q^2E^2(1-\epsilon ))$ and
$\sigma _{\gamma ^*N} = \sigma _T + \epsilon \sigma _L.$
Using
$\sigma_{tot}(\gamma^* N \to \rho N)$ as determined by NMC from
$\mu D $ data,
the  slope of the $t$-dependence of
the cross section, $b = 4.3 \pm 0.6 \pm 0.7 {\rm ~GeV}^{-2}$ as
measured by NMC, and the NMC estimate for
$\sigma_{incoh} /\sigma_{tot} = 0.55 \pm 0.08$ and of
$\sigma^L_{\gamma^* N \rightarrow \rho^0 N}/ \sigma^T_{\gamma^* N
\rightarrow \rho^0 N} \sim 2.0 $ at $Q^2
= 6 {\rm ~GeV}^2$ and $\epsilon=0.8,$
 we can estimate
$${d\sigma^{L\  {\rm experiment}}_{\gamma^* N \rightarrow \rho^0
N}\over dt}\Bigg\vert_{t=0} \sim {14-27{\rm ~nb}\over {\rm GeV}^2}$$
for $Q^2=10 {\rm ~GeV}^2$.
(In deriving this range we also assumed  that
$\sigma_L/\sigma_T$  is either the same
at $Q^2=10~ {\rm GeV}^2$  as at  $Q^2=6~ {\rm GeV}^2$
or increases linearly above $Q^2=6~ {\rm GeV}^2$ as
$\sim Q^2$.)

Let us now compare this result with the leading-logarithmic
prediction of Eq.~\dsigdt. The HMRS $D0'$
parameterization of the gluon distribution in the proton [17]
gives the value $\alpha_s(Q) [xG(x,Q)] = 0.67$
at the NMC kinematics $ x \simeq 0.06,$ $Q^2=10 {\rm
{}~GeV}^2.$
The sensitivity of the parameterization to uncalculated
higher order terms and the uncertainty in the evolution scale
is illustrated by noting that $\alpha_s [xG] = 0.76$ if we use
$\half Q^2$ instead of $Q^2$ for the arguments of $\alpha_s$ and
$G.$ Thus we obtain a range of predictions
$${d\sigma^{L}_{\gamma^* N \rightarrow \rho^0 N}
\over dt}\Big\vert_{t=0}
\sim {(13-17)~- ~(36-47) {\rm ~nb}
\over {\rm GeV}^2}\ .$$
The lower range corresponds to $\eta_{\rho} = 3$
assuming the asymptotic form of the $\rho_L$ distribution amplitude
$\sim z(1-z).$
The upper range corresponds to $\eta_{\rho} = 5$ assuming that the
$\rho_L$ distribution amplitude
is similar to the CZ wave function of a pion.
The  distribution amplitude suggested by Chernyak and Zhitnitski [12]
from QCD sum rules for the $\rho_L$
actually corresponds to a narrower  quark
distribution than for the pion, but it is still
broader than the asymptotic form.
Thus  $\eta_{\rho_L}^{CZ} \sim  3.3 - 3.5$.
If we trust our leading-order estimates, then this value of
$\eta_{\rho_L}$ leads to
$${d\sigma^{L}_{\gamma^* N \rightarrow \rho^0 N}
\over dt}\Big\vert_{t=0, \eta_{\rho_L}^{CZ}} \sim {
16-23{\rm ~nb} \over {\rm GeV}^2}\ ,$$
which is close to the empirical  diffractive $\rho$
leptoproduction cross section.
More generally, $\rho$ electroproduction will allow us to test
determinations of the meson distribution amplitudes from lattice gauge
theory or other non-perturbative QCD computations.

Equation~\dsigdtt\  also allows us to
predict the ratio of the yields of various
vector mesons at $Q^2 \gg m_V^2$ :
$$R(V_1,V_2) \equiv {{d\sigma
(\gamma_L +T \rightarrow V_1 +T)\over dt}\over
{d\sigma (\gamma_L +T \rightarrow V_2 +T) \over dt}}
\bigg\vert_{t=0} = { \Gamma_{V_1}m_{V_1} \over   \Gamma_{V_2}m_{V_2}}
 {\eta_{V_1}^2 \over \eta_{V_2}^2}\ . \eqn\ratio$$
It can be seen from Eq.~\ratio\ that we predict
$R(\phi, \rho)  \sim 1.0 \times e_f^2(\phi)/e_f^2(\rho)$
for $\eta_{\phi}/ \eta_{\rho} \sim 0.9$
suggested by CZ [12]. We also
find  $R(J/\psi, \rho) \sim 1.2
\times e_f^2(J/\Psi)/e_f^2(\rho) $
for $\eta_{J/\psi}=2$ corresponding
to $\phi_{J/\psi} \sim \delta(z-1/2)$.
Here, we have extracted from these
two $R$ results the ratios of the effective
charge squared, where $e_f^2(V)=1/2,1/18,1/9,4/9$ for the
$\rho,\omega,\phi,\Psi$, respectively
(for standard SU(3) wave functions),
which might naively be expected to determine the $R$'s.

Another interesting feature
of the QCD prediction for the production of a vector
meson at large $Q^2$ is the universality of the $t$-dependence of
the process---it is determined
by a universal two-gluon form factor,
independent of the vector meson type.
The upper part of the amplitude corresponding to the
transition $\gamma^*\rightarrow V$ is effectively dipole-like at large
$Q^2;$ \ie\ it should depend weakly on $t$ so long as
$ -t\ll Q^2,$ implying that the $t$-dependence
of the leptoproduction cross section
primarily reflects
the $t$-dependence of the gluon - nucleon scattering amplitude.
The slope of this $t$-dependence should increase
slowly
with the incident energy
due to the Gribov diffusion - shrinkage of the diffractive cone.
The data on exclusive
production of vector  mesons at
high energies supports this prediction---%
the large $Q^2$  $\rho$-meson leptoproduction
cross section has a slope $d\sigma/dt \propto e^{b t}$
corresponding to
$b= 4.3$ GeV$^{-2}$  [14],
which is similar to the small slope  $b \sim  3.5 - 5.5$
GeV$^{-2}$ observed for
exclusive
$J/\psi$-meson photoproduction.

\section{
Non-Perturbative QCD Effects}

The $Q^2$ dependence of the diffractive vector meson leptoproduction
amplitude in the PQCD analysis reflects the overlapping integral
between the light cone wave functions of
$\gamma^*$ and the vector meson. Since
the wave function of $\gamma^*_L$ is $\propto z$ or $(1-z)$ when
$z$ or
$1-z$ vanish, the end-point contributions to the cross section for the
production
of longitudinally polarized vector mesons, arising from Eq.~\mform,
are small and of order $\sim Q^2/ m_V^2(Q^2)^6$.  Since the wave function
of a vector meson bound state should be less singular at $z\sim 0$ and 1
than the wave function of the $\gamma^*$, we conclude that for $\sigma_L$
the end-point contribution may be neglected. Thus  the longitudinal cross
section can be safely calculated in terms of PQCD.

The situation is the opposite in the case of the
production of a vector meson by a  photon with transverse
polarization. The wave function of a transversely polarized
photon is constant at $z\sim 0$ and 1.
As a result, in this case the end-point contribution
is enhanced. To demonstrate the importance of the end-point contribution,
let us consider the process where a final state photon is produced
instead of  a vector meson; \ie\
the process:  $$\gamma^*~+~T~\rightarrow~\gamma~+~T\ .$$
For this reaction, the end-point contribution leads to the cross section
$$ {d\sigma\over dt}^{\gamma^* + T
\rightarrow \gamma + T} \Bigg\vert_{t=0}
\sim {1\over Q^4}\ . \eqn\dsigdtT$$
Now note that the wave function of a transversely polarized
vector meson is less singular than
that of the photon as $ z \rightarrow 0, 1$.
In fact, PQCD predicts that the
$q\bar q$ component decreases at least as fast as
$\propto z$ or $(1-z)$ when $z$ or $1-z$ are small.
Consequently, the cross section for
electroproduction of transversely polarized
svector mesons should fall at least as fast as $1/Q^8$.
(This contribution is additionally suppressed by a double logarithmic
Sudakov-type form factor.)
A similar dependence of the cross section arises from other kinematical
regions.
Thus the experimental investigation of the ratio
of longitudinally and transverse polarized vector
mesons would help to clarify the relative roles of non-perturbative
QCD-end-point contributions and hard physics.

\section{
Diffractive Leptoproduction on Nuclei }

A key feature of the predictions of PQCD for forward
diffractive vector meson leptoproduction
with $1/2m_N x \gg 2R_A$
is the dominance of small size wavefunction configurations.
The fact that the integration
range of $k^2_\perp$ in the vector
meson light-cone wavefunction in Eq.~\phiv\
extends to $\sim Q^2$ implies that the important $q \bar q$
configurations coupling to the virtual photon have transverse separations
$b_\perp \sim 1/Q.$ Thus, even in a nuclear target, color
screening implies that the coherent $q \bar
q$ system can only weakly interact, and in
leading-logarithmic approximation only
two gluons in light-cone gauge connect
the photon-vector meson system to the
nucleus, as illustrated in Fig.~\figB.
Thus, as predicted by PQCD color transparency [18, 11]
the outgoing vector
meson in effect
suffers no final-state absorption, and the
nuclear dependence of the $\gamma^* A \to V A$ forward amplitude will be
identical to that for the case where the final state system is a virtual
photon, $\gamma^* A \to \gamma^* A;$
\ie\ it will be close to additive in the
nucleon number $A$.  We can also understand this remarkable feature of
QCD from the space-time arguments given above: the final state vector
meson is formed from the compact $q \bar q$ pair over a long time
$\tau_f,$ and thus it does not attain its
final physical size and its normal
strong interactions until it is well outside the domain of the target
nucleus. In fact, much of this physics was
anticipated before the advent of
QCD. The possibility that the outgoing absorption of the
$\rho$ in the nucleus would be effectively small in large $Q^2$
leptoproduction was actually first proposed by Yennie in 1975 [19]. The
observation that the incoming photon has
point-like behavior and diminished
absorption was discussed in terms of a
``shrinking photon" by  Cheng and Wu
[20] and by  Bjorken, Kogut, and Soper [21].

We  emphasize that the above reasoning is applicable for a
longitudinally polarized vector meson
only---the effective transverse size of
a produced transversely polarized vector meson is
considerably larger (although still smaller than for ordinary hadrons,
 \cf\ the discussion in Section 2.6).

Although the vector meson suffers no final state interactions, the
forward amplitude ${\cal M}(\gamma^* A \to V A)$ is not strictly additive
in nuclear number since the gluon distribution itself is shadowed.
(This effect is similar to the shadowing of  diffractive production of
high-$p_T$ jets in the $\pi~ +~ A \rightarrow 2 \
\hbox{jets} +A$ reaction,
discussed in Ref.~[8].)
In fact, we see from Eq.~\dsigdt\ that
$${d\sigma\over dt}(\gamma^* A \to V A)\bigg\vert_{t=0} \propto
 \alpha_s^2(Q) [xG_A(x,Q)]^2\ , \eqn\dsigdtrough$$
where $G_A(x,Q)=A^{\alpha_g(x,Q)} G_N(x,Q)$ is the gluon distribution in
the nucleus.  Thus, the analysis presented in this paper
predicts identical nuclear dependence for the forward vector meson
diffractive leptoproduction cross sections, the longitudinal structure
functions $F^L_A(x,Q),$ and the square of the gluon structure functions:
$${{d\sigma\over dt}(\gamma^* A \to V A)\big\vert_{t=0}\over
{d\sigma\over dt}(\gamma^* N \to V N)\big\vert_{t=0}} =
\left [{F^L_A(x,Q) \over F^L_N(x,Q)}\right ]^2
= {G^2_A(x,Q) \over G^2_N(x,Q)} = A^{2 \alpha_g(x,Q)}\ .\eqn\ratios$$
(Note that at finite energies one has to
interpolate the cross section to the unphysical $t=0$ kinematical point.)
The nuclear gluon distribution is expected to be more strongly shadowed
than the nuclear quark structure
functions at intermediate $Q^2$ because of the larger color
charge of the gluon in QCD and thus its stronger inter-nuclear
interactions.  Numerical estimates [22] lead to
$xG_A(x, Q_0^2)/AxG_N(x,Q_0^2) \sim 0.7-0.8~~
(0.4-0.5)$ for $A=12~(200)$ and  $x \sim 0.01-0.03$,  a result
which seems to
be supported by the recent FNAL data of E-665 [23].
However, at fixed $x \sim 0.01-0.03,$
shadowing substantially decreases with
$Q^2$ due to scaling violation effects [22], which should
lead to an effective increase of
transparency for $\rho $ leptoproduction
at fixed $x$ with increasing $Q$.

A nuclear dependence similar to Eq.~\ratios\ is also expected for forward
diffractive $\Upsilon$ and, possibly,
$J/\psi$ leptoproduction cross sections
even at small $Q^2,$ although in this
case the value of $x$ and the evolution
scale of $G(x,Q)$ is controlled by $M_V^2$ rather than $Q^2.$ Thus both
heavy and light diffractive vector meson leptoproduction can provide
basic information on the nature of the gluon distributions in nuclei.

The  calculations given above are applicable
to the  near-forward production of
vector mesons. It should be noted that
the physics relevant to the nuclear dependence
of the leptoproduction cross
section will change with increasing $t$.
We shall give here a semi-quantitative
description of the expected behavior.
At  $-t R_A^2/ 3\ll 1,$
coherent processes dominate the leptoproduction of vector mesons.
The nuclear dependence  of the diffractive
cross section at small $t$  (within
the diffractive peak)
can be estimated  by multiplying Eq.~\ratios\ by the square of
electromagnetic form factor of nucleus
normalized to 1 at $t=0$.
However, if $-t R_A^2/ 3\gg 1$,
incoherent processes, in which the
leading vector meson is accompanied by the production of
other hadrons from nuclear disintegration,
will dominate the cross section.
The existence of nuclear  shadowing implies that
gluons at small $x$ cannot be associated with individual
nucleons.  Thus one can have events
where momenta $-t \ge 0.1~GeV^2 $ are  transferred  to
each of
several  nucleons
which subsequently fragment.
The effect can be a slower $t$-dependence of the cross section and
a smaller energy transfer per interacting
target nucleon than for the scattering
off a single nucleon.
The expected $A$-dependence is intermediate between
that expected for shadowing of $G_A(x,Q^2)$ and $A$.

The recent nuclear target $\rho$ leptoproduction measurements
from the E-665 experiment [16] appear to indicate onset of the color
transparency predicted by PQCD for
incoherent $\gamma^* A \to \rho N(A-1)'$  reactions.
The onset of this phenomena is again for $Q^2 \simeq$ a
few GeV$^2,$ the same scale at which Bjorken scaling is observed in deep
inelastic lepton scattering reactions.
Preliminary
data [15] from the NMC also
confirm  higher
values of the transparency ratio for $Q^2 \ge 3~ {\rm GeV}^2,$
observed in [16]
although the NMC data do not indicate
a $Q^2$ variation of transparency in
their  $10 \ge Q^2 \ge 3 ~{\rm GeV}^2$ range.
One needs to be cautious in interpreting these
data directly in terms of PQCD color transparency of the
outgoing $\rho.$
We note that the
high-$Q^2$ NMC data correspond to
a range of $x$  where the
essential longitudinal distances are smaller
than the nucleus size.  Thus
transparency in this range of kinematics will
reflect to some extent the fact that the
virtual photon can penetrate deeper
into nucleus without interaction.
As we have emphasized,
the nuclear dependence of forward diffractive $\rho$ leptoproduction
which is completely coherent on the
nucleus can provide a decisive test of
color transparency.

\chapter{Conclusions}

The analysis of diffractive leptoproduction cross sections presented in
this paper extends the domain of PQCD predictions to a new domain of
exclusive hadronic reactions.
The central focus of this analysis is
closely related to the calculations of the order $\alpha_s(Q^2)$
leading twist
perturbative contributions to $\sigma_L(\nu, Q^2)$ and the violation of
the Callan-Gross relation.  Although the momentum transfer to the target
is small, the
virtuality of the longitudinally polarized photon
provides
a point-like probe of the diffractive $\gamma^* N \to V N$ process.
At high $Q^2$ the
amplitude factorizes in terms of separable components: the perturbative
distribution
amplitude of the virtual photon, the non-perturbative
distribution amplitude of the
outgoing vector meson system, and a non-perturbative
two-gluon matrix element of the
target closely related to the gluon structure function.  The momentum
transfer dependence of the diffractive amplitude
is thus controlled by a new type of non-local
two-gluon form factor.  We note that since
the momentum transfer to the nucleon is
shared by two gluons, the fall-off can be different from the fall-off
of elastic
electromagnetic form factors.

The factorization analysis can be
extended to the diffractive production of any vector meson system
of mass ${\cal M} $ as long
as $Q^2 > {\cal M}^2.$  The longitudinal cross section always falls as
$1/Q^6$ at fixed ${\cal M},$
$t$ and $s$.  The energy dependence of the forward
diffractive cross section is also universal, reflecting the behavior of
the square of the gluon structure function at $x \sim Q^2/s.$
Related formalisms have also been applied to exclusive heavy quarkonium
photoproduction [4], exclusive pion dissociation to two jets [22],
and to two-photon diffractive reactions [24].

The existing data for diffractive $\rho$ leptoproduction
appear to be consistent even at relatively low $Q^2$ of a
few GeV$^2$ with the leading logarithmic predictions in both magnitude
and in the kinematic dependence.  A crucial feature of the PQCD
predictions, which is clearly evident in the data, is the increasing
dominance of the longitudinal photon and vector meson production
amplitudes with increasing $Q^2.$ More careful studies of the
longitudinal to transverse polarization ratios can lead to insights into
the transition between soft and hard components in QCD amplitudes.
As we have emphasized,
precise measurements of the $A$ dependence of the
diffractive leptoproduction reactions can lead to new insights into
nuclear shadowing of the longitudinal structure function.

In our discussion we have noted only a few of the many empirical tests of
PQCD possible in diffractive leptoproduction.  In principle, the study of
these reactions at HERA will allow tests of the theory over a huge
dynamical range in $x$ and $Q^2.$ The intercomparison of the various
vector meson channels can also lead to new tests and understanding of the
non-perturbative structure of hadronic wavefunctions and their
flavor-symmetry properties.

\bigskip
\centerline{{\bf Acknowledgements}}
\smallskip
We would like to thank
Andrzej Sandacz of NMC and Guang Yin Fang of E-665 for
illuminating discussions of the vector meson leptoproduction data.
SJB, JFG and AM would also like to thank the Institute for Theoretical
Physics, University of California,
Santa Barbara, for support during the early stages of this
project.

\appendix{}
\bigskip

The $\gamma\to V$ transition is defined by
$$\langle 0 \vert J_\mu \vert V \rangle=
{\sqrt 2 ef_V \epsilon^V_\mu\over m_V}
\,.\eqn\fvdef$$
In the frame defined by Eq.~\framedef\
(where we shall take $Q^2=-m_V^2$),
and employing the  time-ordered
formalism outlined in this paper,
we find
$${\sqrt 2 ef_V \epsilon^\gamma\cdot\epsilon^V\over m_V}=
\sqrt{N_c}\sum_{\lambda_1\lambda_2}
\int {d^2k_\perp d^2k^\prime_\perp\over
(16\pi^3)^2}\int dz \int dz^\prime\
\psi^\gamma_{\lambda_1\lambda_2}(k_\perp,z) T \psi^V_{\lambda_1\lambda_2}
(k_\perp^\prime,z^\prime)\,\eqn\basicfv$$
where the amplitude $T$ is trivial,
$$T=16\pi^3\delta(k_\perp^\prime-k_\perp)
\delta(z^\prime-z)\,,\eqn\tform$$
and $\psi^\gamma$ and $\psi^V$ are given by Eqs.~\psigamgeneral\ and
\psivform, respectively.
In the chosen frame, it will be most convenient to isolate $f_V$ by
considering a longitudinally polarized $V$,
with polarization vector as given in Eq.~\polv, while taking
$\epsilon^\gamma_-=1$ and all other components of $\epsilon^\gamma$
to be 0, implying that
$$\epsilon^\gamma\cdot\epsilon^V=\half\epsilon^V_+(L)=
{\half  q_+\over m_V}\,.
\eqn\epsdoteps$$
Meanwhile, in Eq.~\psigamgeneral\ $\gamma\cdot
\epsilon^\gamma=\half\gamma_+$ while in Eq.~\psivform, for $L$
polarization, $\gamma\cdot \epsilon^V$
can be approximated by $\half \gamma_+
\epsilon^V_-=-\half\gamma_+m_V/q_+$. Using the fact that $${\bar
u_{\lambda_1}(k)\over\sqrt z}\ \half\gamma_+
{v_{\lambda_2}(q-k)\over\sqrt{1-z}}
={\bar v_{\lambda_2}(q-k)\over \sqrt{1-z}}\
\half\gamma_+{u_{\lambda_1}(k)\over\sqrt z}
=q_+\delta_{\lambda_1 -\lambda_2}$$
we obtain from Eq.~\basicfv\ the result
(with $Q^2=-m_V^2$ in \psigamgeneral):
$${\sqrt2ef_V(\half q_+)\over m_V^2}
=\sqrt{N_c}ee_fN_V(q_+^2)\sum_\lambda \int
{d^2k_\perp\over 16\pi^3}\int dz\ {\psi^V(k_\perp,z)\over m_V}
\left[{-m_V/q_+\over -m_V^2}\right]\,.\eqn\fvii$$
With $\sum_\lambda=2$ and
using the definition Eq.~\phiv, we obtain
$$\int_0^1\,dz\,\phi^V(z)={f_V\over
\sqrt{N_c}e_f2\sqrt2 N_V}\,.\eqn\integral$$

\endpage

\bigskip
\centerline{{\bf References}}
\smallskip

\pointbegin
 V. N. Gribov, B. L. Ioffe, and I. Ya. Pomeranchuk,
 Yad. Fiz., {\bf 2}, 768
(1965);  B. L. Ioffe,  Phys. Lett. {\bf 30} 123 (1968).
\point
A. Donnachie and P. Landshoff, Phys. Lett. {\bf 185B}, 403 (1987);
Nucl. Phys. {\bf B311}, 509 (1989).
\point
J. R. Cudell, Nucl. Phys. {\bf B336}, 1 (1990).
\point
M. G. Ryskin, Z. Phys. {\bf C57}, 89--92 (1993).
\point
S. J. Brodsky and G. P. Lepage, Phys. Rev. {\bf D22}, 2157 (1980).
\point
B. Z. Kopeliovich, J. Nemchick, N. N.
Nikolaev, and B. G. Zakharov,
KFA-IKP-TH-1993-27, (1993).
\point
B. Bl\"{a}ttel,  G. Baym, L. L. Frankfurt, and M. Strikman,
Phys. Rev. Lett. {\bf 71}, 896 (1993).
\point
L. Frankfurt, G. A. Miller, and M. Strikman, Phys. Lett.
{\bf 304B}, 1 (1993).
\point
S. Catani, M. Ciafaloni, and F. Hautmann,  Phys. Lett. {\bf
B307}, 147--153 (1993).
\point
J. D. Bjorken  in {\it Proceedings of  the International
Symposium on Electron and Photon Interactions at High Energies},
p. 281--297, Cornell (1971);
J. D. Bjorken and J. B. Kogut,  Phys. Rev. {\bf D8}, 1341 (1973).
\point
L. Frankfurt and M. Strikman, Phys. Rept. {\bf 160}, 235 (1988).
\point
V. L.  Chernyak and A. R.  Zhitnitski,  Phys.  Rep.  {\bf 112}, 173
(1984).
\point
EMC, J. J. Aubert et al., Phys. Lett. {\bf 161B}, 203 (1985);
J. Ashman et al., Z. Phys. {\bf C39}, 169 (1988).
\point
NMC, P. Amaudruz et al., Z. Phys. {\bf C54}, 239 (1992).
\point
NMC Collaboration, A. Sandacz, {\it Exclusive $\rho^0$ Muoproduction
at Large $Q^2$},  paper contributed to the poster session
at the {\it Few Body XIV Conference}, Amsterdam, 1993.
\point
Guang Yin Fang, et al., FERMILAB-CONF-93-305, October
1993.
\point
A. D. Martin, W. J. Stirling, and R. G. Roberts,
Phys. Lett. {\bf B306}, 145 (1993).
\point
S. J. Brodsky and A. H. Mueller, Phys. Lett. {\bf 206B}, 685 (1988).
\point
D. R. Yennie, Rev. Mod. Phys. {\bf 47}, 311
(1975).  T. H. Bauer \etal,  Rev. Mod. Phys. {\bf 50}, 261 (1978).
\point
H. Cheng and T. T. Wu, Phys. Rev. {\bf 183}, 1324 (1969).
\point
J. Bjorken, J. Kogut, and D. Soper,
Phys. Rev. {\bf D3}, 1382 (1971).
\point
L. Frankfurt,  M. Strikman, and S. Liuti, Phys. Rev. Lett.
{\bf 65}, 1725 (1990);  Proceeding of PANIC 93 (in press).
\point
Harry L. Melanson, \etal,  FERMILAB-CONF-93-165-E,
Presented at the {\it 28th Rencontres de Moriond: QCD and High
Energy Hadronic
Interactions}, Les Arcs, France, 20--27 March 1993.
\point
I. F. Ginzburg and D. Yu. Ivanov,
Nucl. Phys. {\bf B388}, 376  (1992), and references therein.

\figout

\endpage
\bye